\documentstyle[twocolumn,seceq,graphicx]{jpsj}

\newcommand{\ti}{0}
\newcommand{\tf}{\tau}
\newcommand{\ai}{a_i}
\newcommand{\af}{a_f}
\newcommand{\x}{\mbox{\boldmath$x$}}
\newcommand{\xI}{\x(0)}
\newcommand{\xf}{\x(\tau)}
\newcommand{\diff}{{\mathrm{d}}}

\DeclareRobustCommand{\tmspace}[3]{%
  \ifmmode\mskip#1#2\else\kern#1#3\fi\relax}
\newcommand{\negthickspace}{\tmspace-\thickmuskip{.2777em}}

\def\runtitle{%
Energetics of Open Systems
and
Chemical Potential From Micro-Dynamics Viewpoints
}
\def\runauthor{Tatsuo {\sc Shibata} and Ken {\sc Sekimoto}}

\title{%
Energetics of Open Systems
and
Chemical Potential From Micro-Dynamics Viewpoints
}
\author
{ 
Tatsuo {\sc Shibata}\footnote{E-mail:shibata@kurims.kyoto-u.ac.jp}
and Ken {\sc Sekimoto}$^{1,}$\footnote{E-mail:sekimoto@yukawa.kyoto-u.ac.jp}
}

\inst
{
Research Institute for Mathematical Sciences,
Kyoto University,
Kyoto,
606-8502,
Japan\\
$^1$Yukawa Institute for Theoretical Physics, Kyoto University,
Kyoto 606-8502, Japan
}

\recdate
{
February 25, 2000
}

\abst
{
We present the energetic aspect of open systems
which may exchange particles with their environments.
Our attention shall be paid to the scale that
the motion of the particles is described
by the classical Langevin dynamics.
Along a particular realization of the stochastic process,
we study the energy transfer into the open system
\null from the environments.
We are able to clarify
how much energy each particle carries
when it enters or leaves the system.
On the other hand,
the chemical potential should be considered as 
the concept in macro scale,
which is relevant to
the free energy potential
with respect to the number of particles.
}

\kword
{open systems, stochastic energetics, chemical potential}

\begin{document}
\sloppy
\maketitle

\begin{fullfigure}
\begin{center}
\includegraphics[width=.8\textwidth]{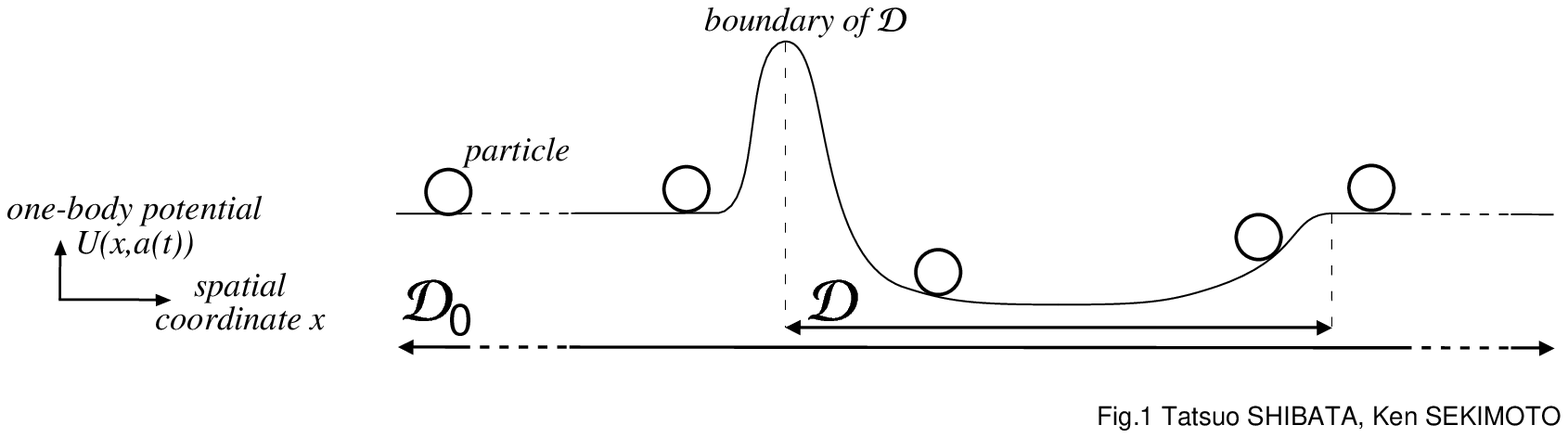}
\end{center}
\caption{%
Schematic diagram of the present situation.
The open system~$\Omega$ consists of the particles
being in the spatial domain~${\cal D}\subset{\cal D}_0$.
Hence, $\Omega$ may exchange particles with its surrounding.
How much energy do the particles carry when these enter or 
leave~$\Omega$?}
\label{fig:}
\end{fullfigure}

\section{Introduction}
\label{sec:Intro}

Open systems may exchange matter with their environments, 
in contact with a heat reservoir.
The energy transfer into such a system \null from its environments
may also take place.
From the viewpoint of energetics,
the question may arise as to
how much energy
the system gains when a particle enters or leaves the system.
In particular,
we may ask the question of
how much energy a particle carries upon migration.
For instance,
we may imagine a macromolecule
which can bind some small molecule,
and may study the variation in the energy of the
macromolecule upon binding~\cite{biomotors}.
According to the thermodynamic relation of open systems:~%
$\Delta E=W+T\Delta S+\mu\Delta n$,
one may expect
that each particle which enters into the system
carries the energy equal to the chemical potential~$\mu$.
%
%
However,
we should notice that
the thermodynamic relation only indicates
the change in the thermodynamic state variables
in a quasi-static process.
In other words,
we can evidently not discuss
the energetic aspect of a particular event of particle exchange,
based on the thermodynamic relation.
Indeed,
we will show that such an argument is not valid.
In this Paper,
the question will be approached
\null from the viewpoint of the microscopic dynamics
which describes
the motion of particles entering or leaving the system.
\null From such a viewpoint,
we will study
the energy balance of open systems upon exchanging a particle.
We shall see that the chemical potential
is obtained as the concepts in the macroscopic level that we find
by considering the ensemble of the dynamics.

In order to study the energy transfer accompanied by
the migration of the particles,
we need a model of the system
which can express the dynamics of 
particles~\cite{Lebowitz,Hill,Julicher}
and 
we need to formulate an energetic interpretation of the model.
In the present Paper,
first we suppose that a closed system in contact with a heat reservoir
consists of~$N$~particles.
Then,  we suppose a subsystem of the closed system.
The subsystem can exchange the particles
with the rest of the closed system.
Thus, we can consider the subsystem as an open system.
By studying the energy transfer of the open system,
we construct the energetics associated with the particle exchange.

Let us first suppose a closed system denoted by~$\Omega_0$
which consists of the $N$ particles.
The dynamics of these particles is assumed to be described properly
by the Langevin dynamics~(here we only consider the overdamped case),
given by
\begin{equation}
\label{eq:NLangevin}
\gamma
{\diff{x_i(t)}\over\diff{t}}=
-{\partial U(\x(t),a(t))\over\partial x_i}
+
\xi_i(t),
\end{equation}
where
$x_i$ is the position of the $i$-th particle,
$U(\x,a)$ is a potential energy
with $\x\equiv(x_1,\cdots,x_N)$ and a controllable parameter~$a$,
$\gamma$ is a friction constant,
and
$\xi_i(t)$ is the white Gaussian processes
which is characterized by
$\langle\xi_i(t)\rangle=0$, and 
$\langle\xi_i(t_1)\xi_j(t_2)\rangle
=2\gamma{\beta^{-1}}\delta_{i,j}\delta(t_1-t_2)$
with $\beta^{-1}=k_BT$.
$k_B$ is the Boltzmann constant
and $T$ is temperature of the heat reservoir.
Throughout this Paper,
we consider the processes 
in which the parameter~$a$ is supposed to be changed
during the time interval \null from $t=0$ to $t=\tau$
by some external system
according to a given protocol~$a(t)$
\null from~$a(0)=\ai$ to $a(\tau)=\af$.

For the energetic interpretation of the Langevin dynamics,
we adopt the {\em stochastic 
energetics}~\cite{Sekimoto1997a,Sekimoto1997b,Sekimoto1998}.
Let us consider
both the friction and fluctuation terms in~Eq.(\ref{eq:NLangevin})
as the force which the heat reservoir exerts on the system.
Then, the work done on the system by the heat reservoir
is given by
\begin{eqnarray}
    \nonumber
    \lefteqn{%
    \sum_{i=1}^N
    \int_{\ti}^{\tf}
    \left(-\gamma{\diff{x_i(t)}\over\diff{t}}+\xi_i(t)\right)\diff{x_i(t)}
    }
    \\
    \nonumber
    &
    =
    &
    \int_{\ti}^{\tf}{\partial U(\x(t),a(t))\over\partial\x}\diff\x(t)
    \\
    &=&
    \int_{\ti}^{\tf}\diff{U(t)}
    -\int_{\ti}^{\tf}{\partial U(\x(t),a(t))\over\partial a}\diff{a(t)},
    \label{eq:heatdef}
\end{eqnarray}
where we use Eq.(\ref{eq:NLangevin})
and 
$\diff{U}=
{\partial U\over\partial a}\diff{a}
+
{\partial U\over\partial \x}\diff\x$.
The left hand side defines
the {\em heat}~$Q[(\xI,\ai){\to}(\xf,\af)]$
along a particular trajectory.
The notation in the bracket~$[\cdot]$
indicates 
that the quantity takes a particular value
which is obtained along 
a particular trajectory~$\x(t)$ and the protocol~$a(t)$.
We should notice that
the above stochastic integral is interpreted as the Stratonovich 
integral~\cite{Gardiner}.
Throughout this Paper,
the stochastic integral is used in the Stratonovich 
sense.

The first term on the right hand side of Eq.(\ref{eq:heatdef})
is the change in the internal energy~$E(t)=U(\x(t),a(t))$,
which we denote by~$\Delta E[(\xI,\ai){\to}(\xf,\af)]$.
The second term is interpreted as 
the work~$W[(\xI,\ai){\to}(\xf,\af)]$
done on the system by some external system
with changing the parameter~$a(t)$ \null from $a(0)=\ai$ to 
$a(\tau)=\af$.
In this way, along one trajectory,
the energy balance relation~\cite{Sekimoto1997a} is obtained as
\begin{eqnarray}
    \nonumber
    \lefteqn{
    \Delta E[(\xI,\ai){\to}(\xf,\af)]
    }\,\,\,\,\,\
    \\
    \nonumber
    &=&
    W[(\xI,\ai){\to}(\xf,\af)]\\
    &&\,\,\,
    +
    Q[(\xI,\ai){\to}(\xf,\af)].
\end{eqnarray}

This Paper has been organized as follows.
A model of the open system will be introduced
in Section~\ref{sec:open}.
The energy balance of open systems
along a particular trajectory is shown in Eq.(\ref{eq:Energy.Consvetion}).
Then we find how much energy each particle carries
at the boundary of the open system.
In Section~\ref{sec:chempot},
it shall be shown that
our study of open system is consistent with thermodynamics.
It will be clarified that
the chemical potential is relevant to the concept of free energy.
Then 
the amount of work onto the open system
in the quasi-static process is discussed
and the thermodynamic relation is obtained.
In Section~\ref{sec:remarks},
we will compare the energy balance relation
with
the thermodynamic relation in Eq.(\ref{eq:quanti}).
Then, we conclude that
the energy which a particle carries upon migration
is not equal to the chemical potential~$\mu$.
We also show some remarks and future problems.

\section{The Energy Balance Relation of Open Systems}
\label{sec:open}

Let us suppose that
the particles of the closed system~$\Omega_0$ are
confined in a spatial domain~${\cal D}_0$,
and
the open system denoted by~$\Omega$
consists of the particles
being in the spatial domain~${\cal D}\subset{\cal D}_0$.
Using some function~$f(x)$,
the spatial domain~${\cal D}$ is specified by $f(x_i)\leq c$.
Hence, $\Omega$ may exchange particles with its surroundings.
See Fig.\ref{fig:} for a schematic illustration of the situation.

We should note the possibility that no particle is in ${\cal D}$.
In such a case,
$\Omega$ contains no degrees of freedom.
In order to avoid such a situation,
it may be easy to introduce
another degrees of freedom into~$\Omega_0$,
that is,
a ``transducer'' which interacts with the particles being in~${\cal D}$.
Then we can define~$\Omega$
which consists of the transducer and the particles 
interacting with the transducer.
However,
it would not provide further physical insight.
In order to present our idea simply,
here we allow the situation that $\Omega$ has no degrees of freedom.

In the following Subsections,
we study the energy balance relation of the open systems.

\subsection{Open systems as a subsystem of a closed system:
One particle case}

Before considering many particle systems,
we present the one particle system, i.e., $N=1$.

Using the Heviside function~$\theta(x)$, 
\begin{equation}
    \theta(x)=
    \left\{
    \begin{array}{ll}
	1&\quad(x\geq0)\\
	0&\quad(x<0)
    \end{array}
    \right.,
\end{equation}
we define
the internal energy~$E(t)$ of~$\Omega$ at time~$t$, given by 
\begin{equation}
    E(t)
    =
    \theta\bigl(c-f(x_1(t))\bigr)
    U(x_1(t),a(t))
\end{equation}

We suppose 
$U(x_1,a)$ depends on~$a$ if $x_1\in{\cal D}$,
i.e.,
$\diff{U(x_1,a)}/\diff{a}=0$ if $x_1\not\in{\cal D}$.
Then,
the change in the internal energy 
during the process is given by
\begin{eqnarray}
    \nonumber
    \lefteqn{
    \Delta E[(x_1(0),\ai){\to}(x_1(\tau),\af)]
    =\int_{\ti}^{\tf}\diff{E(t)}
    }\,\,\,\,\,\,\,\,
    \\
    \nonumber
    &=&
    \int_{\ti}^{\tf}
    \theta\bigl(c-f(x_1(t))\bigr)
    {\partial{U}(x_1(t),a(t))\over\partial a}{\diff{a(t)}}
    \\
    \nonumber
    &&
    +\int_{\ti}^{\tf}
    \theta\bigl(c-f(x_1(t))\bigr)
    {\partial {U}(x_1(t),a(t))\over\partial x_1}\diff{x_i(t)}
    \\
    &&
    +\int_{\ti}^{\tf}
    {\partial\theta\bigl(c-f(x_1(t))\bigr)\over\partial x_1}
    {U}(x_1(t),a(t))\diff{x_1(t)}\,\,\,\,\,.
    \label{eq:dU.open}
\end{eqnarray}

The first term on the right hand side is
the mechanical work~$W[(x_1(0),\ai){\to}(x_1(\tau),\af)]$
on~$\Omega$ done by the external system.
The second term is rewritten by
substituting the Langevin equation~Eq.(\ref{eq:NLangevin}) as
\begin{equation}
    \int_{\ti}^{\tf}
    \theta\bigl(c-f(x_1(t))\bigr)
    \left({-\gamma{\diff{x_1(t)}\over\diff{t}}+\xi_1(t)}\right){\diff{x_1(t)}}.
\end{equation}
Hence, this term is 
the heat~$Q[(x_1(0),\ai)\to(x_1(\tau),\af)]$
transfered into $\Omega$ \null from the heat reservoir.
The third term on the right hand side of~Eq.(\ref{eq:dU.open})
is transformed into
\begin{eqnarray}
    \nonumber
    \lefteqn{
    -\int_{\ti}^{\tf}{\diff{f(x_1)}\over\diff{x_1}}\delta(f(x_1)-c){U}(x_1,a(t))\diff{x_1(t)}
    }
    \\
    \nonumber
    &=&
    -\!\int_{\ti}^{\tf}\!{\diff{f(x_1(t))}\over\diff{t}}%
    \negthickspace\negthickspace\negthickspace\negthickspace
    \negthickspace\negthickspace\negthickspace\negthickspace
    \sum_{\scriptstyle t^{mig}:\atop\scriptstyle f(x_1(t^{mig}))=c}%
    \negthickspace\negthickspace\negthickspace\negthickspace
    {\delta(t-t^{mig})\over\left|{\diff{f(x_1(t^{mig}))}\over\diff{t}}\right|}%
    {U}(x_1(t),a(t))\diff{t}
    \\
    &=&
    \sum_{t=t^{in}}{U}(x_1(t),a(t))
    -\sum_{t=t^{out}}{U}(x_1(t),a(t)),
    \label{eq:1Q^mig}
\end{eqnarray}
where $t^{in}$ and $t^{out}$ indicate all the instants 
that satisfy
$f(x_1(t^{in}))=c,\diff{f(x_1(t^{in}))}/\diff{t}<0$
and $f(x_1(t^{out}))=c,\diff{f(x_1(t^{out}))}/\diff{t}>0$ respectively.
That is, 
$+{U}(x_1,a)$ is added when $f(x_1(t))-c$ varies \null from positive to negative 
and $-{U}(x_1,a)$ is added when $f(x_1(t))-c$ varies \null from negative to positive.
Hence, the value of this term increase or decrease by ${U}(x_1,a)$
when the orbit runs across the boundary of ${\cal D}$ at $x_1$.
We call this term~$Q^{mig}[(x_1(0),\ai){\to}(x_1(\tau),\af)]$. 

Therefore, Eq.(\ref{eq:dU.open}) is rewritten as
\begin{eqnarray}
    \nonumber
    \lefteqn{%
    \Delta E[(x_1(0),\ai)\to(x_1(\tau),\af)] 
    }
    \\
    \nonumber
    &
    =
    &
    W[(x_1(0),\ai){\to}(x_1(\tau),\af)]
    \\
    \nonumber
    &&
    + Q[(x_1(0),\ai){\to}(x_1(\tau),\af)]
    \\
    &&
    + Q^{mig}[(x_1(0),\ai){\to}(x_1(\tau),\af)],
    \label{eq:1energy}
\end{eqnarray}
which displays the energy balance relation of the system.

\subsection{Many particle systems with interactions}
\label{sec:openmany}

Let us next study the many particle systems,
i.e.,
the closed system~$\Omega_0$ contains $N$ interacting particles. 

The interaction potential of the particles 
is the sum of a $n$-body potential for $n=1,2,\cdots$, given by,
\begin{equation}
    \nonumber
    {U}(\x,a)
    =
    \sum_{i_1=1}^N
    \biggl(
    \phi^{(1)}_{i_1}(a)
    +\!\!
    \sum_{i_2=i_1+1}^N
    \Bigl(
    \phi^{(2)}_{i_1,i_2}
    +\cdots
    \Bigr)\biggr)
\end{equation}
where
$\phi^{(n)}_{i_1,\cdots,i_n}\equiv \phi^{(n)}(x_{i_1},\cdots,x_{i_n})$
is the $n$-body energy potential of $x_{i_1},\cdots,x_{i_n}$.
We assume that 
the one body energy potential~$\phi^{(1)}_i(a)\equiv\phi^{(1)}(x_i,a)$
depends on~$a$ if $x_i\in{\cal D}$,
i.e.,
$d\phi^{(1)}(a)/\diff{a}=0$ if $x_i\not\in{\cal D}$.
We assume only short range interactions
so that the $n$ body potential
has non-zero
value only when all the $n$ particles are close to each other.

As in the case of the one particle system,
the internal energy~$E$ of $\Omega$ is defined by 
considering whether each particle is in~${\cal D}$ or not:
\begin{eqnarray}
    \nonumber
    E
    &=&
    \sum_{i_1=1}^N
    \biggl(
    \theta^{(1)}_{i_1}\phi^{(1)}_{i_1}(a)
    +\sum_{i_2=i_1+1}^N
    \Bigl(
    \theta^{(2)}_{i_1,i_2}\phi^{(2)}_{i_1,i_2}
    \\
    &&
    +\sum_{i_3=i_2+1}^N
    \bigl(
    \theta^{(3)}_{i_1,i_2,i_3}\phi^{(3)}_{i_1,i_2,i_3}
    +\cdots
    \bigr)\Bigr)\biggr),
    \label{eq:UOmega}
\end{eqnarray}
where 
$\theta^{(n)}_{i_1,\cdots,i_n}$ is a function
indicating whether the $n$~particles belong to~$\Omega$ or not.
Using the Heviside function~$\theta(x)$,
$\theta^{(n)}_{i_1,\cdots,i_n}\equiv\theta^{(n)}(x_{i_1},\cdots,x_{i_n})$
is defined by
\begin{eqnarray}
    \nonumber
    \theta^{(1)}_{i_1}
    &\equiv&
    \theta(c-f(x_{i_1})),\\
    \theta^{(2)}_{i_1,i_2}
    &\equiv&
    1-(1-\theta^{(1)}_{i_1})(1-\theta^{(1)}_{i_2}),\\
    \nonumber
    \theta^{(n)}_{i_1,\cdots,i_n}
    &\equiv&
    1-(1-\theta^{(1)}_{i_1})(1-\theta^{(1)}_{i_2})\cdots(1-\theta^{(1)}_{i_n}).
\end{eqnarray}
The value of $\theta^{(n)}_{i_1,\cdots,i_n}$ is unity
when at least one of the $n$ particles is in $\Omega$,
or is zero, otherwise.
This means that
certain~$n$~particles belong to~$\Omega$
when at least one particle of these particles is inside~${\cal D}$.
Thus, $\theta^{(n)}_{i_1,\cdots,i_n}=1$
and 
the interaction energy~$\phi^{(n)}_{i_1,\cdots,i_n}$ of 
the~$n$ particles
is included in the internal energy~$E$.
However, 
if $m (<n)$ particles~$j_1,\cdots,j_m$ of these~$n$ particles
are outside~${\cal D}$,
$\theta^{(m)}_{j_1,\cdots,j_m}=0$
and
the interaction energy~$\phi^{(m)}_{j_1,\cdots,j_m}$ of
these~$m$ particles is excluded \null from the 
internal energy~$E$.

The change in the internal energy 
along a particular trajectory is given by,
\begin{eqnarray}
    \label{eq:NDU}
    \nonumber
    \lefteqn{%
    \Delta E[(\xI,\ai){\to}(\xf,\af)]
    }
    \\
    \nonumber
    &=&
    \sum_{i=1}^N\int_{\ti}^{\tf}
    \theta^{(1)}_i(t)
    {\partial\phi_i^{(1)}(a(t))\over\partial a}\diff{a(t)}
    \\
    \nonumber
    &&+
    \sum_{i=1}^N\int_{\ti}^{\tf}
    \theta^{(1)}_{i}(t)
    \left(
    -\gamma{\diff{x_{i}(t)}\over\diff{t}}+\xi_{i}(t)
    \right)\diff{x_{i}(t)}
    \\
    \nonumber
    &&+
    \sum_{i=1}^N\int_{\ti}^{\tf}
    {\partial\theta^{(1)}_i(t)\over\partial x_{i}}\phi_{i}(t)\diff{x_{i}(t)}
    \\
    &&+
    \sum_{i=1}^N\int_{\ti}^{\tf}
    {\partial\phi^{\partial{\cal D}}_{i}(t)\over\partial x_{i}}\diff{x_{i}(t)},
\end{eqnarray}
with
\begin{eqnarray}
    \nonumber
    \phi_{i}
    &=&
    \phi^{(1)}_{i}
    +
    \negthickspace
    \sum_{{\scriptstyle j_1=1\atop\scriptstyle j_1\not=i}}^N
    \negthickspace
    \Bigl(
    \phi^{(2)}_{i,j_1}+
    \negthickspace\negthickspace
    \sum_{{\scriptstyle j_2=j_1+1\atop\scriptstyle j_2\not=i}}^N
    \negthickspace\negthickspace
    \bigl(
    \phi^{(3)}_{i,j_1,j_2}
    +\cdots
    \\
    &&
    \cdots+
    \negthickspace\negthickspace
    \sum_{{\scriptstyle j_{N-1}=j_{N-2}+1\atop\scriptstyle j_{N-1}\not=i}}^N
    \negthickspace\negthickspace
    \phi^{(N)}_{i,j_1,\cdots,j_{N-1}}
    \bigl)
    \Bigl)
    ,
    \label{eq:Ui}
\end{eqnarray}
and
\begin{eqnarray}
    \nonumber
    \phi^{\partial{\cal D}}_{i}
    &=&
    (1-\theta^{(1)}_{i})
    \negthickspace
    \sum_{{\scriptstyle j_1=1\atop\scriptstyle j_1\not=i}}^N
    \negthickspace
    \Bigl(
    \theta^{(1)}_{j_1}\phi^{(2)}_{i,j_1}
    +
    \negthickspace\negthickspace
    \sum_{{\scriptstyle j_2=j_1+1\atop\scriptstyle j_2\not=i}}^N
    \negthickspace\negthickspace
    \bigl(
    \theta^{(2)}_{j_1,j_2}\phi^{(3)}_{i,j_1,j_2}
    +
    \cdots\\
    &&
    \cdots+
    \negthickspace\negthickspace
    \sum_{{\scriptstyle j_{N-1}=j_{N-2}+1\atop\scriptstyle j_{N-1}\not=i}}^N
    \negthickspace\negthickspace
    \theta^{(N-1)}_{j_1,\cdots,j_{N-1}}\phi^{(N)}_{i,j_1,\cdots,j_{N-1}}
    \bigr)
    \Bigr)
    .
    \label{eq:Uouti}
\end{eqnarray}

Eq.(\ref{eq:NDU}) is obtained by the following transformation.
The total derivative of the internal energy~Eq.(\ref{eq:UOmega}) is
written as
\begin{eqnarray}
    \nonumber
    \diff{E}
    &=&
    \sum_{i=1}^N
    \theta_i^{(1)}
    {\partial\phi_i(a)\over\partial a}\diff{a}
    \\
    \nonumber
    &&+
    \sum_{i=1}^N
    {\partial\over\partial x_i}
    \biggl(
    \sum_{j_1=1}^N
    \Bigl(
    \theta^{(1)}_{j_1}\phi^{(1)}_{j_1}
    +
    \negthickspace\negthickspace\negthickspace
    \sum_{j_2=j_1+1}^N
    \negthickspace\negthickspace\negthickspace
    \bigl(
    \theta^{(2)}_{j_1,j_2}\phi^{(2)}_{j_1,j_2}
    +\cdots
    \\
    &&
    \cdots
    +\sum_{j_N=j_{N-1}+1}^N
    \theta^{(N)}_{j_1,\cdots,j_N}\phi^{(N)}_{j_1,\cdots,j_N}
    \bigr)
    \Bigr)
    \biggr)\diff{x_i}
    \label{eq:NDU0}
\end{eqnarray}
Considering that 
$\theta^{(N)}_{i_1,\cdots,i_N}$ and $\phi^{(N)}_{i_1,\cdots,i_N}$
are symmetric functions of $x_{i_1},\cdots,x_{i_N}$,
the second term on the right hand side of~Eq.(\ref{eq:NDU0})
is rewritten as
\begin{eqnarray}
    \nonumber
    &&
    \sum_{i=1}^N
    \negthickspace
    {\partial\over\partial x_{i}}
    \negthickspace
    \biggl(
    \theta^{(1)}_{i}\phi^{(1)}_{i}
    \!+
    \negthickspace
    \sum_{\scriptstyle j_1=1\atop\scriptstyle j_1\not=i}^N
    \negthickspace
    \Bigl(
    \theta^{(2)}_{i,j_1}\phi^{(2)}_{i,j_1}
    \!+
    \negthickspace\negthickspace\negthickspace
    \sum_{\scriptstyle j_2=j_{1}+1\atop\scriptstyle j_2\not=i}^N
    \negthickspace\negthickspace\negthickspace
    \bigl(
    \theta^{(3)}_{i,j_1\!,j_2}\phi^{(3)}_{i,j_1\!,j_2}
    \negthickspace
    +\!\cdots
    \\
    &&
    \cdots+
    \negthickspace\negthickspace\negthickspace
    \sum_{\scriptstyle j_{N-1}=j_{N-2}+1\atop\scriptstyle j_{N-1}\not=i}^N
    \negthickspace\negthickspace\negthickspace
    \theta^{(N)}_{i,j_1,\cdots,j_{N-1}}\phi^{(N)}_{i,j_1,\cdots,j_{N-1}}
    \bigr)
    \Bigr)
    \biggr)
    \diff{x_i}
    \label{eq:2nd}
\end{eqnarray}
Noting that $\theta^{(n)}_{i_1,\cdots,i_n}$ is
transformed as 
$
\theta^{(n)}_{i_1,\cdots,i_n}
=\theta^{(1)}_{i_1}+(1-\theta^{(1)}_{i_1})
\theta^{(n-1)}_{i_2,\cdots,i_n},
$
then Eq.(\ref{eq:2nd}) is rewritten as
\begin{equation}
    \sum_{i=1}^N
    \biggl(
    \theta^{(1)}_{i}
    {\partial\phi_{i}\over\partial x_{i}}
    +
    {\partial\theta^{(1)}_{i}\over\partial x_{i}}
    \phi_{i}
    +
    {\partial\phi^{\partial{\cal D}}_{i}\over\partial x_{i}}
    \biggr)
    \diff{x_i},
    \label{eq:2nd2}
\end{equation}
with Eqs.(\ref{eq:Ui})~and~(\ref{eq:Uouti}).
Using the relation
$
    {\partial\phi_{i}\over\partial x_{i}}
    =-\gamma{\diff{x_{i}}\over\diff{t}}+\xi_{i},
$
Eq.(\ref{eq:2nd2}) is rewritten as
\begin{equation}
    \sum_{i=1}^N
    \biggl(
    \theta^{(1)}_{i}
    \Bigl(%
    {-\gamma{\diff{x_{i}}\over\diff{t}}+\xi_{i}}%
    \Bigr)
    +
    {\partial\theta^{(1)}_{i}\over\partial x_{i}}
    \phi_{i}
    +
    {\partial\phi^{\partial{\cal D}}_{i}\over\partial x_{i}}
    \biggr)\diff{x_i}.
    \label{eq:2nd3}
\end{equation}
Substituting
Eq.(\ref{eq:2nd3}) into Eq.(\ref{eq:NDU0}),
and integrating it with respect to 
$\x(t)$ and $a(t)$ \null from $t=\ti$ to $t=\tf$,
then a simple calculation leads to~Eq.(\ref{eq:NDU}).

The first term on the right hand side of Eq.(\ref{eq:NDU})
is the work~$W[(\xI,\ai){\to}(\xf,\af)]$
on~$\Omega$ done by the external system.
The second term of Eq.(\ref{eq:NDU}) is
the heat transfered \null from the heat reservoir
to the particles being in~${\cal D}$,
denote by~$Q[(\xI,\ai){\to}(\xf,\af)]$.
As is discussed in the previous Section,
the third term of Eq.(\ref{eq:NDU})
equals to $\phi_i(t^{in})$
when the orbit~$x_{i}(t)$
enters into~${\cal D}$ over the boundary at~$t=t^{in}$,
or equals to  $-\phi_i(t^{out})$ when the orbit~$x_{i}(t)$
goes out of~${\cal D}$ at $t=t^{out}$.
Let us denote
this energy transfer accompanied by the migration of the particles
by $Q^{mig}[(\xI,\ai){\to}(\xf,\af)]$.
The fourth term of Eq.(\ref{eq:NDU})
distinguishes 
the interacting particles systems
\null from one particle systems or
free particles systems.
This term shows
the contribution by the particles outside~${\cal D}$
which interacts with the particles inside~${\cal D}$
across the boundary~$\partial{\cal D}$ of~${\cal D}$.
Here, we denote this term as~$Q^{\partial{\cal D}}[(\xI,\ai){\to}(\xf,\af)]$.

Therefore, Eq.(\ref{eq:NDU}) is rewritten as
\begin{eqnarray}
    \nonumber
    \lefteqn{%
    \Delta E[(\xI,\ai){\to}(\xf,\af)]
    }\,\,\,\,\,
    \\
    \nonumber
    &=&
    W[(\xI,\ai){\to}(\xf,\af)]
    \\
    \nonumber
    &&
    + Q[(\xI,\ai){\to}(\xf,\af)]
    \\
    \nonumber
    &&
    + Q^{mig}[(\xI,\ai){\to}(\xf,\af)]
    \\
    &&
    + Q^{\partial{\cal D}}[(\xI,\ai){\to}(\xf,\af)]
    \label{eq:Energy.Consvetion}
\end{eqnarray}
In this way,
along a particular trajectory,
we have the energy balance relation of the open systems
with many body interactions.

In the similar way,
the change in
the ensemble average of the internal energy~$E(t)$
can be also discussed.
The ensemble average of the internal energy is defined by
\begin{equation}
\langle E(t)\rangle=\int E(t)P(\x,t)\diff\x,
\label{eq:Ensemble.U}
\end{equation}
where $P(\x,t)$
is the probability distribution function of the state of~$\Omega_0$
obtained by solving the Fokker-Plank equation,
and $\langle\cdot\rangle$ indicates the ensemble average
over all the possible trajectory.

The energy balance relation of the mean values is 
\begin{eqnarray}
    \nonumber
    \Delta\langle E[\ai{\to}\af]\rangle
    =
    \langle W[\ai{\to}\af]\rangle
    +\langle Q[\ai{\to}\af]\rangle
    \\
    +
    \langle Q^{mig}[\ai{\to}\af]\rangle
    +\langle Q^{\partial{\cal D}}[\ai{\to}\af]\rangle.
    \label{eq:Ensemble.thermo}
\end{eqnarray}
Henceforth we use the notation~$[\ai{\to}\af]$
for those quantities
concerning the processes in which~$a(t)$ is changed \null from $\ai{\to}\af$.
Each term on the right hand side
corresponds to each term in Eq.(\ref{eq:Energy.Consvetion}).
We do not need to postulate further condition
in order to obtain this expression for the mean values.
Comparing Eq.(\ref{eq:Energy.Consvetion})
with the thermodynamic relation
that we have presented in Section~\ref{sec:Intro},
we have
\begin{equation}
    \langle Q\rangle
    +
    \langle Q^{mig}\rangle
    +
    \langle Q^{\partial{\cal D}}\rangle
    =
    T\Delta S+\mu\Delta n,
    \label{eq:heats}
\end{equation}
where we suppose the quasi-static process.
We will discuss the meaning of this expression
in Section~\ref{sec:remarks}.

Before concluding this Section,
let us evaluate~$Q^{mig}$
for the case that we are able to effectively exclude 
the possibility of the direct interaction 
{\em across} the boundary~$\partial{\cal D}$,
i.e.~$Q^{\partial{\cal D}}=0$~\cite{situation}.
We further suppose~${U}(\x,a)=\tilde{U}(\x,a)$~defined by
\begin{equation}
    \tilde{U}(\x,a)\equiv\tilde{U}^{(n)}(x_1,\cdots,x_n,a)+(N-n)U_0,
    \label{eq:Phi}
\end{equation}
when $x_i$ satisfies $f(x_i)<c$,
$(i=1,\cdots,n)$
and $x_j$ satisfies $f(x_j)\geq c$,
$(j=n+1,\cdots,N)$.
This assumption leads to the simple form of~$Q^{mig}$
as 
\begin{equation}
    Q^{mig}
    =
    U_0
    \Bigl(
    \sum_{t=t^{in}}1-\sum_{t=t^{out}}1
    \Bigr).
\end{equation}
Even in the present simple case,
the relation between the terms on the left and the right 
hand side of Eq.(\ref{eq:heats})
does not seem evident.
However,
this does {\em not} imply that
our study of the open system
based on the stochastic energetics
is {\em in}compatible with thermodynamics.
In the next Section,
we will show the consistency.
Then in Section~\ref{sec:remarks},
Eq.(\ref{eq:heats})
is consistently interpreted
mentioning the difference of the scale of description
on each term.

\section{%
The thermodynamic relation \null from microscopic dynamics}
\label{sec:chempot}

It has been shown in the previous 
works~\cite{Sekimoto1997a,Sekimoto1997b,Sekimoto1998}
that
the framework of the stochastic energetics
is consistent with thermodynamics
of the system in contact with a heat reservoir.
In this Section,
we shall show that
even in the open system
the stochastic energetics is consistent with thermodynamics~\cite{Sasa1998}.
First, in Subsection~\ref{subsec:A}
we study the probability
that the open system~$\Omega$ is found to be
a particular physical state
which is specified by
the number of particles in ${\cal D}$
with their positions.
Then,
we will find that
the chemical potential
is relevant to the concept of free energy potential
obtained through the reduction of the description of the system.
In Subsection~\ref{subsec:B},
we also study the quasi-static process,
in which the work onto the open system is shown to equal
the difference of a pertinent thermodynamic potential.
Then, in Subsection~\ref{subsec:C}
the thermodynamic relation is obtained
so that the compatibility of our study with thermodynamics shall be shown.

In the present Section,
the state of the closed system~$\Omega_0$
is specified by~$\x^{(N)}=(x_1,\cdots,x_N)$
which constitute the phase space~$\Gamma^{(N)}$.
We assume that all the particles are identical.
If two points in $\Gamma^{(N)}$
differ only by the permutation of $x_i$'s
they may be considered as representing the same {\em physical state}.
Then,
we can introduce the phase space~$\Gamma^{\prime(N)}$
obtained by the identification of all such points.
In $\Gamma^{\prime(N)}$
the state of the particles
is specified by $\x^{\prime(N)}$.
For each point $\x^{\prime(N)}$ in ${\Gamma}^{\prime(N)}$,
there are $N!$ points in $\Gamma^{(N)}$.
Hence,
if $f(\x^{(N)})$ is a symmetric function of all the $x_i$'s
then
\begin{equation}
    \int_{
    {{({\cal D}_0}^N)}^\prime}
    f(\x^{\prime(N)})\diff\x^{\prime(N)}
    =
    {1\over N!}
    \int_{
    {{\cal D}_0}^N}
    f(\x^{(N)})\diff\x^{(N)},
    \label{eq:int}
\end{equation}
where $({{\cal D}_0}^N)'$ indicates the domain in 
${\Gamma}^{\prime(N)}$
which corresponds to ${{\cal D}_0}^N$ in $\Gamma^{(N)}$.

\subsection{The chemical potential as free energy potential}
\label{subsec:A}

The equilibrium distribution function of~$\x^{(N)}$~of~$\Omega_0$ is 
obtained by solving the Fokker-Plank equation,
\begin{equation}
    P_{eq}(\x^{(N)},a)
    =
    {1\over{\cal Z}(a)}{1\over 
    v_0^{N}}
    e^{-\beta U(\x^{(N)},a)}.
\end{equation}
with the normalization condition:
\begin{equation}
    1=
    {1\over v_0^N}
    {1\over{\cal Z}(a)}
    \int_{({{\cal D}_0}^N)}%
    e^{-\beta U(\x^{(N)},a)}\diff\x^{(N)}.
\end{equation}
Here
the constant~$v_0$ of the dimension of volume has been introduced
so that ${\cal Z}(a)$ is dimensionless.
Since ${U}(\x^{(N)},a)$  is symmetric function of $x_i$'s,
the probability distribution function of the state $\x^{\prime(N)}$ is 
\begin{equation}
    P_{eq}^{\prime}(\x^{\prime(N)},a)
    =
    {N!\over{\cal Z}(a)v_0^{N}}
    e^{-\beta U(\x^{\prime(N)},a)},
\end{equation}
with $\x^{\prime(N)}\in ({\cal D}_0^N)^\prime.$

By integrating 
$P_{eq}^{\prime}(\x^{\prime(N)},a)$
with respect to the degrees of freedom of the particles 
being outside~${\cal D}$, that is, in ${\cal D}_0\backslash{\cal D}$,
we obtain 
the probability distribution function of~$\Omega$ that
there are $n$ particles in~${\cal D}$
with their positions~$\x^{\prime(n)}\in({{\cal D}}^n)^\prime$ as
\begin{equation}
    P_{eq}^{\prime}(\x^{\prime(n)},a)
    =
    {1\over{\cal Z}(a)}{1\over v_0^{n}}
    e^{-\beta{\cal U}^{(n)}(\x^{\prime(n)},a)},
\end{equation}
where ${\cal U}^{(n)}(\x^{\prime(n)})$ is considered as the free energy potential
under the restriction that the state of $\Omega$ is $\x^{\prime(n)}$.
Hence, ${\cal U}^{(n)}(\x^{\prime(n)})$ is given by
\begin{equation}
    e^{-\beta{\cal U}^{(n)}\!(\x^{\prime(n)},a)}
    =
    {N!\over v_0^{N-n}}
    \int_{%
    (({\cal D}_0\backslash{\cal D})^{N-n})^\prime}
    \negthickspace\negthickspace\negthickspace\negthickspace
    \negthickspace\negthickspace\negthickspace\negthickspace
    e^{-\beta U\!(\x^{\prime(N)},a)}
    \diff\x^{\prime(N-n)},
    \label{eq:eff.potential}
\end{equation}
with $\x^{\prime(n)}\in({{\cal D}}^n)^\prime$.
We can show the following normalization condition holds:
\begin{equation}
    1
    =
    \sum_{n=0}^N
    \int_{({{\cal D}^{n})^\prime}}\!\!\!%
    P_{eq}^{\prime(n)}(\x^{\prime(n)},a)
    \diff\x^{\prime(n)}.
\label{eq:ppeqnorm}
\end{equation}

Hereafter we shall exclude the possibility 
of the direct interaction
among the particles across the boundary~$\partial{\cal D}$
and assume 
$U(\x^{(N)},a)=\tilde{U}(\x^{(N)},a)$, with Eq.(\ref{eq:Phi}).
Then, Eq.(\ref{eq:eff.potential}) is rewritten as
\begin{eqnarray}
    \nonumber
    \lefteqn{
    e^{-\beta{\cal U}^{(n)}\!(\x^{\prime(n)},a)}
    }
    \\
    &=&
    {N!(V_0-V)^{N-n}\over\!v_0^{N-n}(N-n)!}%
    e^{-\beta\bigl(\tilde{{U}}^{(n)}\!(\x^{\prime(n)},a)
    +(N-n) U_0\bigr)},
\end{eqnarray}
with $\x^{\prime(n)}\in({{\cal D}}^n)^\prime$.
We shall consider the case that 
the volume of $\Omega_0$,~$V_{0}\equiv\int_{{\cal D}_0}\!1\diff{x}$ and $N$
are sufficiently large with $\rho\equiv N/V_0$ and 
$V\equiv\int_{{\cal D}}1\diff{x}$ being kept to be finite constants.
Then, ${\cal U}^{(n)}(\x^{\prime(n)},a)$ is given by
\begin{eqnarray}
    \nonumber
    \lefteqn{
    {\cal U}^{(n)}(\x_{n},a)
    =
    \tilde{{U}}^{(n)}(\x^{\prime(n)},a)
    -n\bigl(U_0+\beta^{-1}\log{\rho v_0}\bigr)
    }
    \\
    &&
    +N\Bigl(U_0+\beta^{-1}\log{v_0\over V_0}\Bigr)
    +\beta^{-1}\rho V
    +{\cal O}\Bigl({1\over N}\Bigr).
\end{eqnarray}
Since, on the right hand side, only
the first and the second terms 
depend on $\x^{\prime(n)}$ and on $n$,
these terms
should be regarded as a new free energy potential
assigned to the situation where
there are $n$ particles in $\cal D$
with their positions
corresponding to
the physical state~$\x^{\prime(n)} \in ({\cal D}^n)^\prime$.
Hence,
introducing
the {\em chemical potential}~$\mu$ through the definition:
\begin{equation}
    \mu\equiv U_0+\beta^{-1}\log{\rho v_0},
    \label{eq:chem.pot}
\end{equation}
${P_{eq}^{\prime(n)}}(\x^{\prime(n)},a)$ is rewritten as
\begin{equation}
    {P_{eq}^{\prime(n)}}(\x^{\prime(n)},a) 
    =
    {1\over\Xi(a)}
    {1\over v_0^{n}}
    e^{-\beta
    \bigl(
    \tilde{{U}}^{(n)}(\x^{\prime(n)},a)
    -\mu n
    \bigr)}, 
\label{eq:pdf.omega}
\end{equation}
where $\Xi(a)$ is a normalization constant given by
\begin{eqnarray}
    \nonumber
    \Xi(a)
    &=&
    \lim_{N\to \infty}
    {{\cal Z}(a)}
    e^{\beta{N
    \bigl(
    U_0+\beta^{-1}\log{v_0\over V_{0}}
    +{\cal O}\left(1\over N\right)
    \bigr)}}
    \\
    &=&
    \sum_{n=0}^\infty
    {1\over v_0^{n}}
    \int_{({\cal D}^n)^\prime}\!\!\!%
    e^{-\beta
    \left(
    \tilde{{U}}^{(n)}(\x^{\prime(n)},a)
    -\mu n
    \right)}
    \diff\x^{\prime(n)}\,\,\,\,\,.
\end{eqnarray}
Eq.(\ref{eq:pdf.omega}) is the {\em grand canonical distribution}.

Notice that 
the chemical potential~$\mu$
is obtained as a part of 
the free energy potential~${\cal U}^{(n)}(x_{1},\cdots,x_{n},a)$.
Hence,
the crucial point for obtaining the chemical potential
is to consider the trajectory~$\x^{\prime(N)}$
in the reduced space $\Gamma^{\prime(N)}$ rather than $\Gamma^{(N)}$ 
{\it and} to extract the effective potential of the state~$\x^{\prime(n)}
\in({\cal D}^n)^\prime$ 
by eliminating the degrees of freedom 
in~$(({\cal D}_0\backslash{\cal D})^{N-n})^\prime$.

\subsection{Work on the open system in quasi-static process}
\label{subsec:B}

As is discussed in the previous Section,
the work onto the open systems
is a random variable,
whose value is determined
depending on a particular trajectory.
If the process is quasi-static, however,
we expect that 
the amount of work is 
deterministic quantity.
In this Subsection,
we show that the work onto the open system is given by
the difference of a pertinent thermodynamic potential.

Here we note that
the Langevin dynamics~Eq.(\ref{eq:NLangevin})
has the ergodic property
that long time averages 
can be replaced by the ensemble averages.
When the change of the parameter~$a$ is slow enough
and hence $\left|\diff{a(t)}/\diff{t}\right|$ is small enough,
the work along a particular trajectory
given by
$
\int_{0}^\tau
{\partial U(\x^{\prime(N)}(t),a(t))\over\partial a}
\diff{a(t)}
$
can be evaluated asymptotically as the
integral of
the equilibrium ensemble average of the integrand 
with respect to $a$ \null from $a_i$ to $a_f$.
Using ${P_{eq}^{\prime(n)}}(\x^{\prime(n)},a)$,
the work done by the external system
is written in the limit~$\tf\to\infty$ as
\begin{eqnarray}
    \nonumber
    \lefteqn{
    \lim_{\tau{\to}\infty}
    W[(\x(0),\ai){\to}(\x(\tau),\af)]
    }
    \\
    \nonumber
    &=&
    \int_{\ai}^{\af}\!\!\diff{a}
    \sum_{n=0}^N
    \int\limits_{({\cal D}^n)^\prime}
    \negthickspace
    \diff\x^{\prime(n)}
    {\partial\tilde{U}(\x^{\prime(n)},a(t))\over\partial a}
    {P_{eq}^{\prime(n)}}(\x^{\prime(n)},a)
    \\
    &=&
    \Delta J[\ai{\to}\af]
    =
    J(\af)-J(\ai),
    \label{eq:work.open}
\end{eqnarray}
where $J(a)$ is the {\em grand canonical potential}  defined by
$J(a)=-\beta^{-1}\log{\Xi(a)}.$

Hence,
if the change of the parameter is slow enough,
the work done by the external system
onto the open system 
equals to the difference of the grand canonical potential
with the probability being unity,
without mentioning the ensemble average.

It should be noted that
for a finite time interval,
the amount of the work of a particular process can be larger or smaller 
than~$\Delta J$.
Considering the ensemble average of the work,
the minimum work principle holds
on the open system~\cite{Sasa1998},
i.e.,
$\langle W[\ai{\to}\af]\rangle
\geq\Delta J[\ai{\to}\af]$.

\subsection{Thermodynamic relation}
\label{subsec:C}

The equilibrium ensemble average of the internal energy of~$\Omega$ is given by
\begin{eqnarray}
    \nonumber
    \lefteqn{%
    \langle E(a)\rangle
    =
    \sum_{n=0}^N
    \int_{({\cal D}^n)^\prime}\!\!\!%
    \tilde{{U}}^{(n)}(\x^{\prime(n)})
    P_{eq}^{\prime(n)}(\x^{\prime(n)},a)
    \diff\x^{\prime(n)}
    }\,\,\,
    \\
    \nonumber
    &&=
    \sum_{n=0}^N%
    \!%
    \int_{({\cal D}^n)^\prime}%
    \negthickspace
    \Bigl(
    -\beta^{-1}\!\log{%
    \bigl(%
    v_0^{n}P_{eq}^{\prime(n)}(\x^{\prime(n)},a)%
    \bigr)}%
    +\mu n+J(a)%
    \Bigr)%
    \\
    &&
    \,\,\,\,\,\,\,\,\,\,\,\,\,\,\,\,\,\,\,\,\,\,
    \times
    P_{eq}^{\prime(n)}(\x^{\prime(n)},a)
    \diff\x^{\prime(n)},
\end{eqnarray}
where we substitute~Eq.(\ref{eq:pdf.omega}) into $\tilde{U}^{(n)}(\x^{\prime(n)},a)$. 
Introducing the {\em entropy}~$S(a)$ given by
\begin{equation}
    S(a)
    =
    -k_B\sum_{n=0}^N
    \int_{({\cal D}^n)^\prime}\!\!\!%
    P_{eq}^{\prime(n)}
    \log{\bigl(
    v_0^{n}P_{eq}^{\prime(n)}
    \bigr)}
    \diff\x^{\prime(n)},
\end{equation}
we have
\begin{equation}
\langle E(a)\rangle
\
=J(a)+TS(a)+\mu\langle n(a)\rangle,
\end{equation}
where $\langle n(a)\rangle$ is
the equilibrium ensemble average of the number of particles in~${\cal D}$.

When the parameter~$a$ is changed \null from~$\ai$~to~$\af$
quasi-statically,
the change in the internal energy is given by,
\begin{eqnarray}
    \nonumber
    \lefteqn{
    \Delta\langle E[\ai\to\af]\rangle
    =
    W[\ai\to\af]
    }\,\,\,\,\,\,\,\,\,\,\,\,
    \\
    &&+T\Delta S[\ai\to\af]
    +\mu\Delta\langle n[\ai\to\af]\rangle,
    \label{eq:Thermo}
\end{eqnarray}
where Eq.(\ref{eq:work.open}) is substituted.
In this way, we have the thermodynamic relation of the open system,
which holds on the ensemble of the trajectories~$\x(t)$.

Eq.(\ref{eq:Thermo}) may suggest that the increase in number of 
particles during a quasi-static process
would accompany the transfer of energy into the system~$\Omega$ 
as if each particle carries the chemical potential.
In the next Section,
we conclude this Paper,
mentioning how the thermodynamic relation is compatible 
with the energy balance relation 
that we have obtained in Eq.(\ref{eq:Energy.Consvetion}).

\section{The energy carried by a particle upon migration: Conclusion}
\label{sec:remarks}

In the preceding sections,
we have constructed the energy balance relation~%
Eq.(\ref{eq:Energy.Consvetion}) of the open system
focusing our attention to the scale
that the system is described by the Langevin dynamics.
On the other hand,
it has been clarified that
the chemical potential,
which reflects the probability that
the open system contains a particular number of particles,
is  relevant to the free energy potential.
Here, we shall illustrate
the compatibility of the energy balance relation~Eq.(\ref{eq:Energy.Consvetion})
with the thermodynamic relation~Eq.(\ref{eq:Thermo}),
noting the difference of the scale of their description.

As is the case in the previous Section,
here we shall consider the case that 
the possibility of the direct interaction
among the particles across~$\partial{\cal D}$
is effectively excluded,
and we suppose the quasi-static limit and $V, N{\to}\infty$. 
\null From Eqs.(\ref{eq:Energy.Consvetion}) and (\ref{eq:Thermo})
we have the quantitative relation
\begin{equation}
    \langle Q\rangle
    +
    \langle Q^{mig}\rangle
    =
    T\Delta S+\mu\Delta\langle n\rangle,
    \label{eq:quanti}
\end{equation}
and $Q^{\partial{\cal D}}=0$.
If we assume
$U(\x,a)=\tilde{U}(\x,a)$ with Eq.(\ref{eq:Phi}), 
we have $Q^{mig}=U_0\Delta n.$
Remembering the definition of~$\mu$ given by Eq.(\ref{eq:chem.pot}),
we have,
\begin{equation}
    \langle Q^{mig}\rangle\not=\mu\Delta\langle n\rangle.
\label{eq:inequality}
\end{equation}
This leads to our main conclusion of the present Paper that 
{\em each particle does not carry~$\mu$
across the boundary of the system,
when it enters or leaves the system.}

We should notice
the difference between the energy balance relation associated with the
left hand side of Eq.(\ref{eq:quanti})
and the thermodynamic relation associated with the
right hand side of Eq.(\ref{eq:quanti}).
The thermodynamic relation is a closed expression in terms of the
variation in the thermodynamic state variables of a system.
On the other hand,
the energy balance relation
is the relation of energy exchange with the surroundings.
Let us study
how $\langle Q\rangle$ is expressed by
using the thermodynamic state variables.
\null From Eq.(\ref{eq:quanti}) and $Q^{mig}=U_0\Delta n$,
we have
\begin{equation}
    \langle Q\rangle
    =
    T\Delta S
    +
    k_{B}T\log{\bigl(\rho v_0\bigr)}
    \Delta\langle n\rangle
    \not=T\Delta S.
    \label{eq:heat.open}
\end{equation}
This quantitative relation shows that
the ensemble average of
the energy transfer \null from the heat bath
is expressed by the variation in 
the two thermodynamic state variables~$\Delta S$ and $\Delta\langle n\rangle$
in both the two terms~$T\Delta S$ and $\mu\Delta\langle n\rangle$ of the thermodynamic relation.

We should also mention
the difference of the level of the description between the energy balance
relation and the thermodynamics relation.
Since the energy balance relation holds on a particular trajectory,
it is not necessary for obtaining the relation
to know the probability of the realization of the trajectory.
In this sense, 
the energy balance relation describes the microscopic scale.
On the other hand,
in order to obtain the thermodynamic relation,
it is necessary to know the probability distribution of the physical states.
The thermodynamic relation is obtained so that it is a relation among 
the ensemble averages.
In this sense,
the thermodynamic relation describes the macroscopic level.
Consequently, the chemical potential is regarded as
the concept in the macroscopic description.

Thus,
when we observe a particular process, i.e. a sample path,
we can not find the chemical potential
as well as the thermodynamic relation,
but measure a particular value of~$Q$ obeying the energy balance relation.
With the increase of the number of the samples,
many different values of~$Q$ shall be recorded.
Then, we know that $Q$ takes a value following some probability distribution.
Eventually,
we find the chemical potential and the thermodynamic relation
by examining the ensemble of the processes.

It may be a future problem
to study an energy conversion cycle
of a micro-machine with at least two particle reservoirs.
For example, suppose that 
the two reservoirs consist of the particles of the same species
and provide the same one-body energy potential
and the same temperature environment.
Then the difference of the chemical potentials 
$\mu_H$ and $\mu_L$ of these reservoirs can arise only through 
the difference of the density of the particles.
Thermodynamics tells that the maximally available 
mechanical work obtained per particle transported is 
$\mu_H-\mu_L$~$(>0)$~\cite{Julicher,Shibata1998}.
By studying this process with the micro-machine,
it might be more evident
how the small fluctuating system
{\em feels} the chemical potential so as to produce the mechanical work.



\begin{thebibliography}{99}
\bibitem{biomotors}
    Biological processes of biopolymers should  
    be approached from the viewpoint of the energetics of open systems.
    The theoretical studies have been appeared in 
    Refs.\citen{Hill,Julicher},
    where, however, the energetics of  a single event of particle 
    exchange has been discussed little.
\bibitem{situation}
    We may consider a dilute distribution of hard-core particles in the 
    surroundings~(${\cal D}_0\backslash{\cal D}$).
\end{thebibliography}
\end{document}